\documentclass{article}
\usepackage{graphicx}
\usepackage[english]{babel}
\usepackage[margin=1in]{geometry}
\usepackage[ruled]{algorithm2e}
\usepackage{multicol}
\usepackage{subcaption}
\usepackage[square,sort,comma,numbers]{natbib}

\author{S. Chatterjee \footnote{\texttt{rs171@iiita.ac.in}} \and B.S. Sanjeev \footnote{\texttt{sanjeev@iiita.ac.in}} }
\title{
{\bfseries\Large Network strategies to study Epstein-Barr virus associated carcinomas and potential etiological mechanisms for oncogenesis\bigskip}
}

\date{July 31, 2022}

\begin{document}
\maketitle

\begin{center}
Department of Applied Sciences\\ Indian Institute of Information Technology\\ Allahabad 211012, India \\
\end{center}


%

\tableofcontents


\begin{abstract}

Diseased conditions are a consequence of some abnormality that are associated with clinical conditions in numerous cells and tissues affecting various organs. The common role of EBV (Epstein-Barr virus) in causing infectious mononucleosis (IM) affecting B-cells and epithelial cells and the development of EBV-associated cancers has been an area of active research. Investigating such significant interactions may help discover new therapeutic targets for certain EBV-associated lymphoproliferative (Burkitt's Lymphoma and Hodgkin's Lymphoma) and non-lymphoproliferative diseases (Gastric cancer and Nasopharyngeal cancer). Based on the DisGeNET (v7.0) data set, we constructed a disease-gene network bipartite graph to identify genes that are involved in various carcinomas namely, gastric cancer (GC), nasopharyngeal cancer (NPC), Hodgkin's lymphoma (HL) and Burkitt's lymphoma (BL). Using the community detection algorithm (Louvain method), we identified communities from the disease-gene network followed by functional enrichment using over-representation analysis methodology to detect significant biological processes/pathways and the interactions between them. There is increasing evidence that genes responsible for causing a particular disease are not distributed randomly but in localized communities with some common functional properties. In this study, we identified the modular communities to explore the relation of this common causative pathogen (EBV) with different carcinomas such as GC, NPC, HL and BL. We could identify the top 10 genes as CASP10, BRAF, NFKBIA, IFNA2, GSTP1, CSF3, GATA3, UBR5, AXIN2 and POLE based on their degree of distribution. Further over-representation analysis showed that the ABL1 gene was significantly over-represented in 3 out of 9 critical biological processes, namely in regulating pathways in cancer, the TP53 network and the Imatinib and chronic myeloid leukemia biological processes. As a result, we can infer that the EBV pathogen is selective in targeting critical pathways to bring about cellular growth arrest/apoptosis and interfering with vital biological processes, including the TP53 network of genes that leads to further proliferation of damage to vital cellular activities.


\end{abstract}

\section{Introduction}
\label{intro}

Epstein-Barr virus (EBV) affects the vast majority (more than 90\%) of the global population. It was the first virus to be involved in the carcinogenesis of lymphoma and in the development of virus-associated lymphoid and epithelial tumors \citep{epstein1964virus}. Evidence suggests that EBV mainly affects B-lymphocytes, but it can also affect NK-cells, T-cells and epithelial cells \citep{hue2020epstein}. It is a member of the herpes virus family and spreads through body fluids, most commonly through saliva. It has been an area of research that the co-existence of EBV infection in many cancer patients provides the scope for a sustained role of EBV in developing infectious mononucleosis (IM) and EBV-associated cancer development. It is also of potential interest to note that the presence of EBV in almost all the malignant cells provides a strong correlation between EBV and associated cancers \citep{babcock1998ebv}.

EBV has a higher incidence in causing infectious mononucleosis (IM) \citep{henle1968relation} and increases the risk of developing several malignancies mainly associated with gastric cancer (GC), nasopharyngeal cancer (NPC), Hodgkin's lymphoma (HL) and Burkitt's lymphoma (BL). The precise role involved in the molecular mechanisms involved in the etiology of EBV-associated cancers like GC, NPC, HL and BL are still not yet fully understood \citep{shannon2017epstein}. As there are no medicines or vaccines present to prevent the infection caused by EBV, it is of vital interest to study gene-level perspectives involved in the pathogenesis of developing these malignancies \citep{cohen2018vaccine}. Hence, it becomes vital to investigate various processes involved in viral oncogenesis as such insights would ultimately contribute to a better understanding of tumor pathogenesis to reduce the chances of the severity of the infection and reduce EBV-associated malignancies by developing targeted cancer therapies. 

Previous research studies have revealed the effects of EBV oncogenes, immunosuppression and the exploitation of the immune modulators such as B-cells and NK-cells that contributes to the development of EBV-associated carcinomas \citep{murata2014modes}. Studies have documented that through EBV infection, there could be a possibility of a sustained contribution to the development of carcinomas which explores the mechanisms that are thought to be important in cancers associated with Epstein-Barr virus, such as GC, NPC, HL and BL. In this study, through a network-based integrated approach, we attempt to decipher the mechanisms of oncogenesis and the critical processes involved in each of the EBV-associated cancer. The objective of our study is to implement a topological network analysis of the genes involved in the association among GC, NPC, HL and BL to discover significant biological processes and the inter-connected modular communities between them for the development of targeted therapeutic outcomes for clinical studies.

Interactions between genes and proteins are indispensable to fundamental cellular processes as they perform various functions. Network analysis tools using graph-theoretic approaches enable network topological information to analyze functional dependencies between various disease-gene interactions linked to Epstein-Barr virus associated carcinomas \citep{wang2021identification}. The emergence of Epstein-Barr virus associated diseases is of critical interest to various carcinomas (mainly GC, NPC, HL and BL) and other syndromes of clinical significance \citep{farrell2019epstein}. Insights from such interactions between EBV-associated cancers can be beneficial to understand the intrinsic molecular mechanisms. Given the global incidence of such carcinomas, it is vital to investigate the significance of molecular interactions in patients affected with such known neoplastic conditions.

Network centrality approaches have been previously modeled to study such molecular level interactions (PPIs). This network-based approach has been effectively used to understand various disease mechanisms. In the current study, we constructed a bipartite graph of the disease-gene network, which maps 22 EBV-associated carcinomas broadly categorized into 4 major classes (GC, NPC, HL and BL) involving 382 human genes (see Figure 1\label{EBV_PPI}).

\begin{figure}
\centering
\includegraphics[width=6.5in]{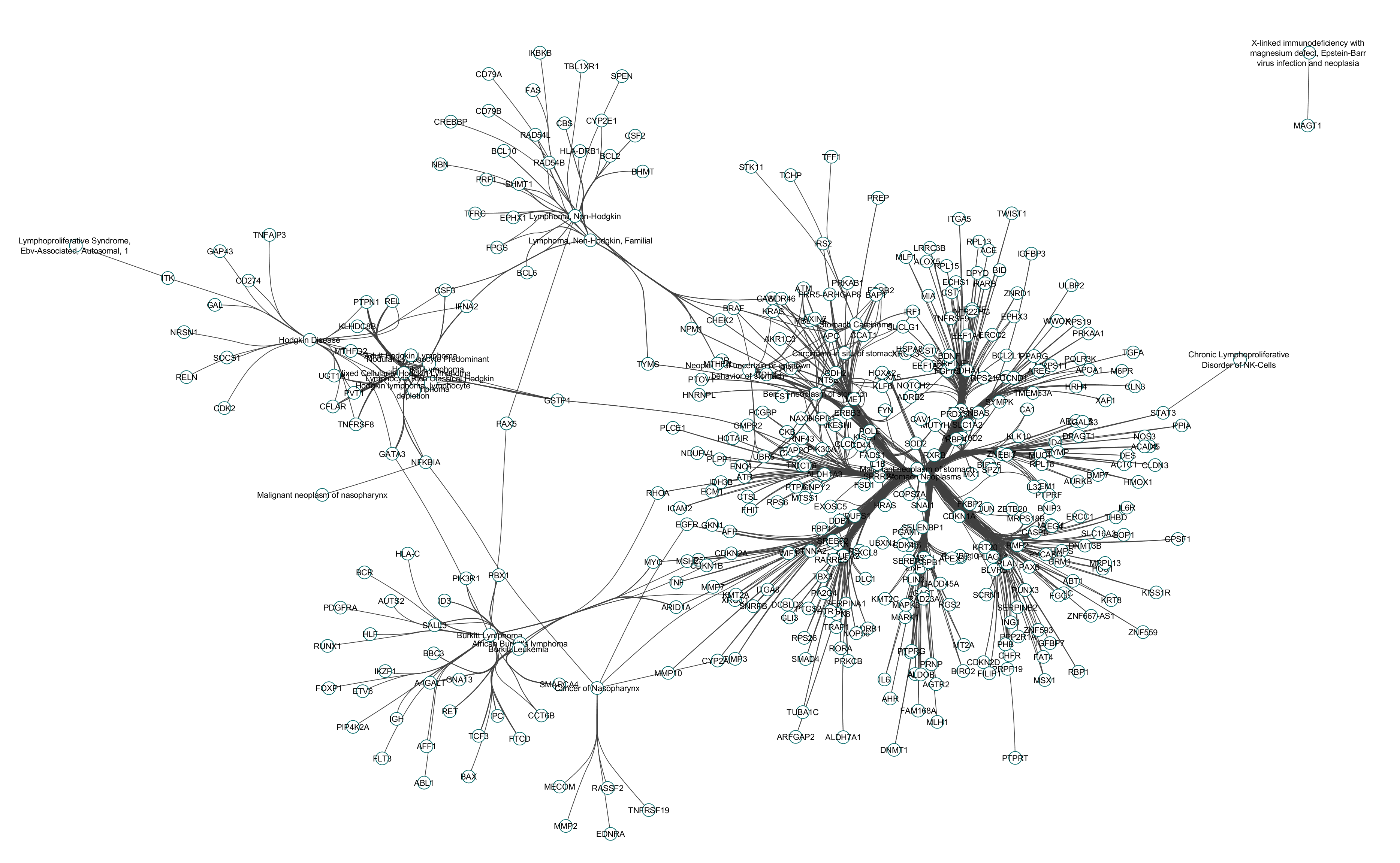} 
\caption{The Disease Gene Network of Epstein Barr Virus in various neoplastic processes (Gastric cancer (GC), Nasopharyngeal cancer (NPC), Hodgkin's lymphoma (HL) and Burkitt Lymphoma (BL)}.
\label{EBV_PPI}
\end{figure}

\begin{table*}[!t]
\centering
\caption{{Types of cancer associated with Epstein-Barr virus.}\label{table:Host_Pathogen_Interactome_Data}}
{\begin{tabular}{llll}
\hline
Sl.	& Category & Associated Cancer Type & Reference [PMID]\\
\hline
1 & & Malignant neoplasm of stomach						& 1342957 \\
2 & & Stomach Neoplasms									& 1342957 \\
3 & GC & Stomach Carcinoma								& 1342957 \\
4 & & Benign neoplasm of stomach							& 1342957 \\
5 & & Carcinoma in situ of stomach						& 1342957 \\
6 & & Neoplasm of uncertain / unknown behavior of stomach	& 1342957 \\
\hline
7 & & Lymphoma, Non-Hodgkin								& 9547991 \\
8 & & Hodgkin Disease									& 9547991 \\
9 & & Lymphoma, Non-Hodgkin, Familial					& 9547991 \\
10 & & Adult Hodgkin Lymphoma							& 9547991 \\
11 & HL & Hodgkin lymphoma, lymphocyte depletion 			& 9547991 \\
12 & & Lymphocyte Rich Classical Hodgkin Lymphoma			& 10505767 \\
13 & & Mixed Cellularity Hodgkin Lymphoma					& 10505767 \\
14 & & Nodular Lymphocyte Predominant Hodgkin Lymphoma 	& 10505767 \\
\hline
15 & & Burkitt Lymphoma									& 2824192 \\
16 & BL & Burkitt Leukemia 								& 2824192 \\
17 & & African Burkitt's lymphoma						& 2824192 \\
\hline
18 & NPC & Cancer of Nasopharynx							& 8732865 \\
19 & & Malignant neoplasm of nasopharynx					& 8732865 \\
\hline
20 & & Lymphoproliferative Syndrome, Ebv-Associated, Autosomal	& 22409825 \\
21 & Syndrome & Chronic Lymphoproliferative Disorder of NK-Cells 	& 6158759 \\
22 & & X-linked immunodeficiency with magnesium defect	& 25313976 \\
\hline
\end{tabular}}
\end{table*}

\section{Materials and Methods}
\label{sec:1}

The methodology uses network theoretic tools for community detection and over-representation analysis for functional enrichment.

\subsection{Disease-gene Network of EBV linked to GC, NPC, HL and BL}
\vspace{0.1cm}
DisGeNET (v7.0) is a comprehensive database containing collections of genes and variants associated with human diseases. DisGeNET (v7.0) integrates data from expert curated repositories, GWAS databases, animal models and the scientific literature \citep{pinero2020disgenet}. This curated data set contains disease-associated molecular interactions involving human genes and the interactions are derived from multiple databases. It includes interaction data using high throughput technologies from multi-level proteomics studies.

We constructed the disease-gene network linked to Epstein-Barr virus (EBV) from previously known referenced \citep{thompson2004epstein} sources (see Table 1) using the DisGeNET (v7.0) data set as an undirected bipartite graph $G$ ($V_G$, $E_G$), where the nodes ($V_G$) represent genes and diseases, and edges ($E_G$) represent disease-gene interactions. This biparted graph comprises disease-gene interactions that map genes with the corresponding associated diseases linked with Epstein-Barr virus (EBV). Based on evidence-based clinical research studies, we filtered the disease-gene interactions that interact with the Epstein-Barr virus.

\subsection{Network Node degree distribution}
\vspace{0.1cm}
The degree of a node is defined as the number of edges connected to a node. It is one of the basic parameters to study the network topology of a graph and is primarily used to study network connectivity in graphs as one of the centrality measures. The number of certain high-degree nodes in the network defines the scale-free distribution of hubs in the overall network. It is one of the most intuitive centrality measures used in protein-protein interaction network \citep{jeong2001lethality}. Figure 2 \label{N_C_D} shows the number of nodes (EBV-associated diseases) corresponding to the number of degrees. Malignant neoplasm of stomach (in GC category) was the most affected carcinoma with maximum number of genes (300) linked to it as shown in Table 2.\label{N_C_D}. The select top 10 nodes in all EBV-associated carcinomas out of total 382 unique genes with respect to their degrees were identified as CASP10, BRAF, NFKBIA, IFNA2, GSTP1, CSF3, GATA3, UBR5, AXIN2 and POLE. Consequently, we performed functional gene enrichment analysis using over-representation methodology as it revealed critical pathways that could well be targeted as they were over-represented among the gene sets.

\begin{figure}[h!]
\centering
\includegraphics[width=3.0in]{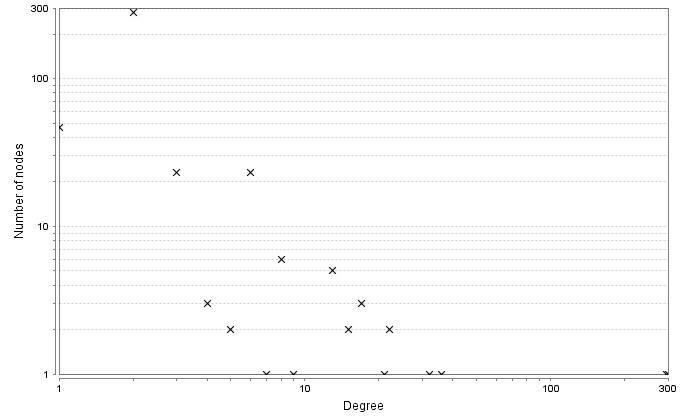}
\caption{The node degree distribution of the EBV-associated cancers as depicted in Table 2. Malignant neoplasm of stomach (in GC category) was the most affected carcinoma associated with EBV.}
\label{N_C_D}
\end{figure}

\begin{table*}[!t]
\centering
\caption{{Number of diseases and their classes along with their degree of distribution.}\label{table:Host_Pathogen_Interactome_Data}}
{\begin{tabular}{llll}
\hline
Sl.	& Disease & Class	& Degree \\
\hline
1 & Malignant neoplasm of stomach						& Neoplastic Process 	& 300 \\
2 & Stomach Neoplasms									& Neoplastic Process 	& 297 \\
3 & Stomach Carcinoma									& Neoplastic Process 	& 21 \\
4 & Benign neoplasm of stomach							& Neoplastic Process 	& 17 \\
5 & Carcinoma in situ of stomach							& Neoplastic Process 	& 17 \\
6 & Neoplasm of uncertain / unknown behavior of stomach	& Neoplastic Process 	& 17 \\
\hline
7 & Lymphoma, Non-Hodgkin								& Neoplastic Process 	& 32 \\
8 & Hodgkin Disease										& Neoplastic Process  & 22 \\
9 & Lymphoma, Non-Hodgkin, Familial						& Neoplastic Process 	& 22 \\
10 & Adult Hodgkin Lymphoma								& Neoplastic Process 	& 13 \\
11 & Hodgkin lymphoma, lymphocyte depletion 				& Neoplastic Process	& 13 \\
12 & Lymphocyte Rich Classical Hodgkin Lymphoma			& Neoplastic Process 	& 13 \\
13 & Mixed Cellularity Hodgkin Lymphoma					& Neoplastic Process	& 13 \\
14 & Nodular Lymphocyte Predominant Hodgkin Lymphoma 		& Neoplastic Process 	& 13 \\
\hline
15 & Burkitt Lymphoma									& Neoplastic Process  & 36 \\
16 & Burkitt Leukemia 									& Neoplastic Process  & 15 \\
17 & African Burkitt's lymphoma							& Neoplastic Process  & 15 \\
\hline
18 & Cancer of Nasopharynx						 		& Neoplastic Process 	& 9 \\
19 & Malignant neoplasm of nasopharynx					& Neoplastic Process 	& 1 \\
\hline
20 & Lymphoproliferative Syndrome, Ebv-Associated, Autosomal	& Disease / Syndrome 	& 1 \\
21 & Chronic Lymphoproliferative Disorder of NK-Cells 	& Neoplastic Process 	& 1 \\
22 & X-linked immunodeficiency with magnesium defect	& Disease / Syndrome 	& 1 \\
\hline
\end{tabular}}
\end{table*}

\subsection{Network Modularity in EBV-linked Carcinoma Network}

Modularity is defined as the fraction of edges that fall within the given groups minus the expected fraction, if edges were distributed at random \citep{brandes2008robert}. A positive modularity signifies that the number of edges connected to nodes in a network within a community exceeds the expected value by chance randomly. To identify significant communities Louvain method-based community detection was implemented, with p-values representing the significance \citep{sham2014statistical}. The network modularity of the disease-gene network was found to be 0.359 which depicted a positive modular network structure (see Figure 3\label{N_M}).

\begin{figure}[h!]
\centering
\includegraphics[width=3.0in]{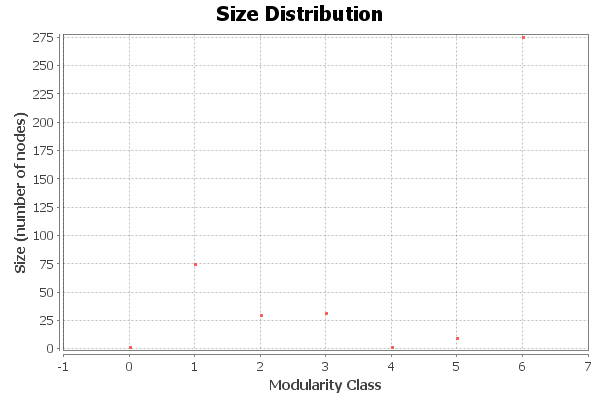}
\caption{The network modularity of the disease-gene network in EBV-associated carcinomas was 0.359 signifying a positive structured modular network. The figure depicts the size distribution of the 7 modularity classes.}
\label{N_M}
\end{figure}

\subsection{Community detection in modular networks}
\vspace{0.1cm}

Community detection is fundamental to biological network analysis for protein function annotation, disease gene prediction and studies based on targeted drug therapies \citep{singhal2020multiscale}. Such techniques have successfully been implemented for the identification of critical genes based on modularity, network modules and drug targets, leading to clinically significant outcomes \citep{gulbahce2008art}. Community detection methods build hierarchical representations directly from the graph structure of the EBV-affected carcinoma disease-gene network. Based on the hierarchical organization of the network, it assigns nodes to communities. We implemented Community Detection Application and Service framework (CDAPS) based on the integrated approach \citep{singhal2020multiscale} to identify and visualize large-scale multi-scale network communities. In order to detect hierarchical communities, we used Hierarchical community Decoding Framework (HiDeF), which is based on multi-scale approach for network community detection in biological networks \citep{zheng2021hidef}.

We performed community detection using the Louvain algorithm in the CDAPS framework with a maximum resolution parameter of less than 50 to gain insights even from relatively smaller communities. The community persistence threshold was kept at 5, to remove unstable clusters. This resulted in a hierarchical network with communities as nodes and their hierarchical relationships as edges. We fetched significant communities in the form of hierarchical modules in the overall network. For functional enrichment, the communities were enriched based on over-representation analysis within the gene sets from curated databases for gene ontology. The Louvain method \citep{blondel2008fast} based on network modularity (Q) is defined as:

\begin{eqnarray}
{\displaystyle Q={\frac {1}{2m}}\sum \limits _{ij}{\bigg [}A_{ij}-{\frac {k_{i}k_{j}}{2m}}{\bigg ]}\delta (c_{i},c_{j})}
\end{eqnarray}

where ${\displaystyle A_{ij}}$ represents the edge weight between nodes ${\displaystyle i}$ and ${\displaystyle j}$;
${\displaystyle k_{i}}$ and ${\displaystyle k_{j}}$ are the sum of the weights of the edges attached to nodes ${\displaystyle i}$ and ${\displaystyle j}$ respectively; ${\displaystyle m}$ is the sum of all of the edge weights in the graph; ${\displaystyle c_{i}}$ and ${\displaystyle c_{j}}$ are the communities of the nodes; and ${\displaystyle \delta }$  is Kronecker delta function $( {\displaystyle \delta (x,y)=1}$ if ${\displaystyle x=y}$, ${\displaystyle 0}$ otherwise).

\begin{algorithm}[h!]
\bigskip
\SetAlgoLined
\SetKwRepeat{REPEAT}{repeat}{until}
\SetKwFor{IF}{if}{then}{endif}
\SetKwFor{FOR}{for}{do}{endfor}
\SetKwFor{WHILE}{while}{do}{endwhile}
G the initial network\\
\REPEAT{}{
{Put each node of G in its own community}\;
\WHILE {some nodes are moved}{
\FOR {all node n of G}{
place n in its neighboring community\\ including its own which maximizes\\ the gain in modularity
}
}
\IF {the new modularity is greater than the initial}{
G = the network between communities of G\;
\Else{Terminate}}
}
\caption{Louvain Algorithm for Community Detection}
\label{L_A}
\end{algorithm}

\subsection{Over-representation Enrichment Analysis}

We employed the methodology of over-representation analysis to detect statistically significant gene ontology terms. To detect functional information, we used g:Profiler that underlines biological processes with known functional information and associations from Ensembl database enriched in the gene lists \citep{raudvere2019g}. Using g:Profiler, genes were mapped to known functional annotations to detect statistically significant enriched pathways for functional enrichment (see Tab. 4\label{tbl:biological_process}). This process allows over-representation of cellular-level information that highlights statistically significant pathways.

The nodes were selected with minimum Jaccard index value \citep{jaccard1912distribution} for overlap greater than 0.05 and the range of p-value \citep{storey2003statistical} was kept less than 0.00001. Databases such as the KEGG database \citep{kanehisa2000kegg}, WikiPathways \citep{pico2008wikipathways} and the Gene Ontology database (GO) \citep{gene2004gene} were used to fetch relevant information for gene ontology in various biological processes.
 
\begin{figure}[h!]
	\centering
	\includegraphics[width=6.5in]{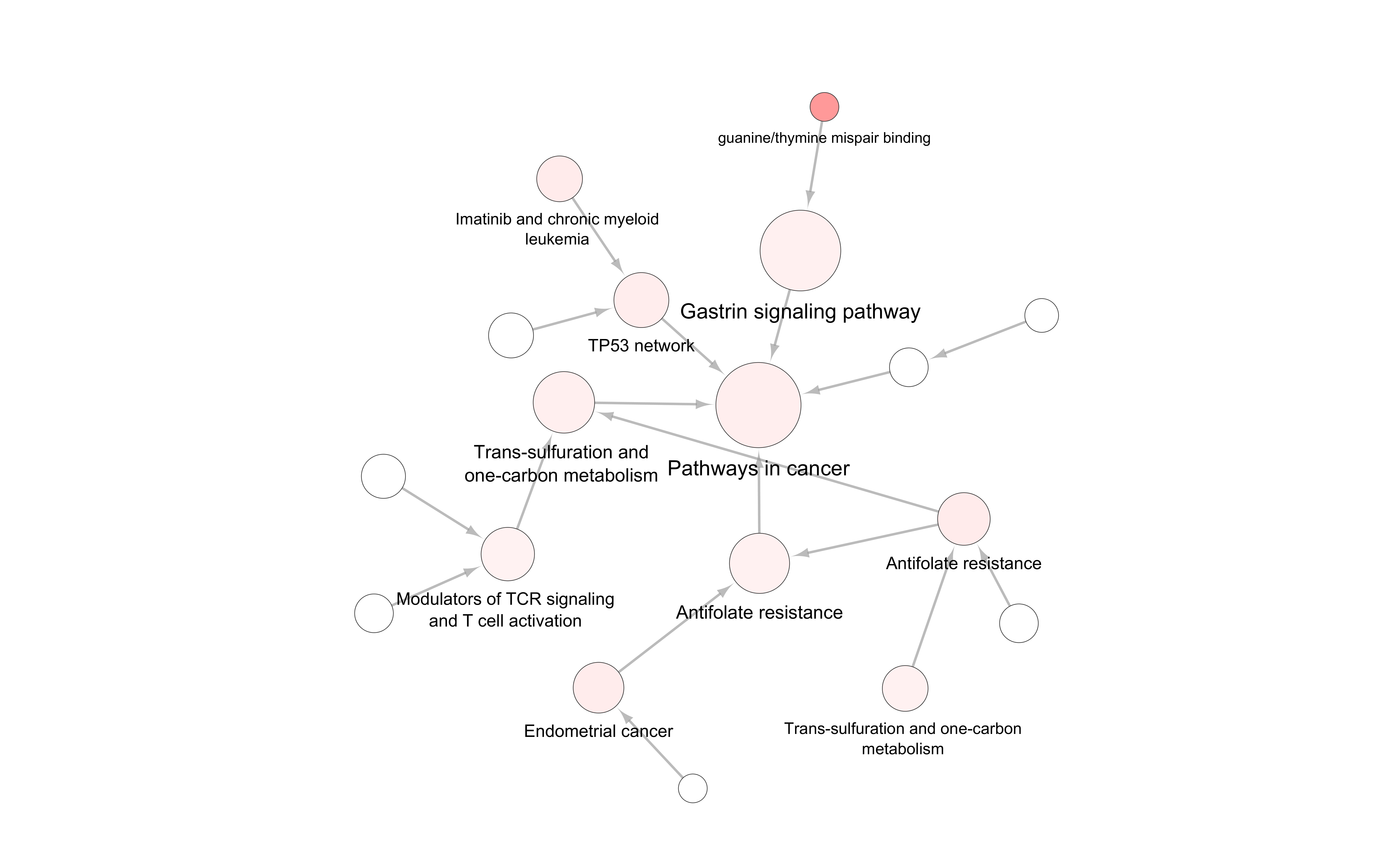}
	\caption{The figure depicts nodes in the Disease-Gene Network which represents its significance denoted by the size of the node (based on p-values). The respective biological processes and pathways are interlinked in a hierarchical network layout using over-representation analysis. The Hierarchical community Decoding Framework was used and Louvain method was implemented to detect statistically significant communities.}
	\label{N_C_D}
\end{figure}

\begin{table*}[!ht]
\centering
\caption{{Significant biological processes and the list of genes involved in each community process based on their degree of distribution.}
\label{table:Gene_Pathogen_Dataset}}
{\begin{tabular}{p{0.2in} p{2in} p{0.3in} p{3.0in}}
\hline
Sl.No & Biological process & Degree & Genes \\
\hline
1 & Pathways in cancer & 70 & ABL1 APC AXIN2 BAX BBC3 BCL2 BCL2L1 BCR BID BIRC2 BIRC5 BMP2 BRAF CASP8 CCND1 CDH1 CDK2 CDK4 CDKN1A CDKN1B CDKN2A CREBBP CTNNA2 CXCL8 EDNRA EGFR ERBB2 F2R FAS FGFR2 FLT3 GADD45A GLI3 GNA13 GSTP1 HMOX1 HRAS IFNA2 IKBKB IL6 IL6R JUN KRAS MAPK1 MAPK3 MAPK8 MECOM MET MLH1 MMP2 MSH2 MYC NFKBIA NOTCH2 PDGFRA PIK3CA PIK3R1 PPARG PRKCB PTGS2 RARB RET RHOA RUNX1 RXRB SMAD4 STAT3 TGFA TP53 TPM3 \\
\hline
2 & Gastrin signaling pathway & 27 & BCL2L1 BIRC2 BIRC5 BMP2 CCND1 CD44 CDKN1A CDKN1B CDKN2A CXCL8 EGFR FYN GAST HRAS JUN KRAS MAPK1 MAPK3 MAPK8 MMP7 PPARG PTGS2 RHOA RPS6 SERPINB2 SERPINE1 STAT3 \\
\hline
3 & Endometrial cancer & 7 & APC AXIN2 BRAF CDH1 ERBB2 FGFR2 PIK3CA \\
\hline
4 & Trans-sulfuration and one carbon metabolism & 6 & BHMT CBS MTHFD2 MTHFR SHMT1 TYMS \\
\hline
5 & Antifolate resistance & 5 & FPGS IKBKB MTHFR SHMT1 TYMS\\
\hline
6 & Modulators of TCR signaling and T-cell activation & 5 & ITK NFKBIA REL SOCS1 TNFAIP3\\
\hline
7 & TP53 Network & 4 & ABL1 BAX BBC3 MYC\\
\hline
8 & Imatinib and chronic myeloid leukemia & 3 & ABL1 BCR PDGFRA\\
\hline
9 & Guanine / Thymine mispair binding & 2 & MLH1 MSH2\\
\hline
\end{tabular}}
\end{table*}

\begin{table*}
	\small
	\caption{List of significantly enriched biological processes and pathways along with their corresponding p-values (threshold less than 2.43E-06).}
	\label{tbl:biological_process}
	\begin{tabular*}{\textwidth}{@{\extracolsep{\fill}}llllll}
		\hline
		Rank & Gene Ontology / Biological process & P-value & -log(P-value) & Reference \\
		\hline
		1 & Pathways in Cancer & 4.78E-69 & 68.32 & KEGG:05200\\
		2 & Gastrin signaling pathway & 8.58E-37 & 36.06 & WP:WP4659\\
		3 & Endometrial cancer & 3.30E-15 & 14.48 & WP:WP4155\\
		4 & Trans-sulfuration and one carbon metabolism & 1.58E-11 & 10.80 & WP:WP2525\\
		5 & Antifolate resistance & 6.82E-11 & 10.17 & KEGG:01523\\
		6 & Modulators of TCR signaling and T-cell activation & 1.79E-09 & 8.75 & WP:WP5072\\
		7 & TP53 Network & 1.23E-08 & 7.91 & WP:WP1742\\
		8 & Imatinib and chronic myeloid leukemia & 3.86E-07 & 6.41 & WP:WP3640\\
		9 & Guanine / Thymine mispair binding & 2.43E-06 & 5.61 & GO:0032137\\
		\hline
	\end{tabular*}
\end{table*}

\begin{table*}[h!]
	\small
	\caption{Validated of functionally enriched biological processes using STRING along with the total number of genes in the community (G$_c$) and the False Discovery Rates (FDR).}
	\label{tbl:fdr}
	\begin{tabular*}{\textwidth}{@{\extracolsep{\fill}}lllll}
		\hline
		Rank & Gene Ontology / Biological process & G$_c$ & FDR \\
		\hline
		1 & Pathways in Cancer & 70 & 2.32E-103\\
		2 & Gastrin signaling pathway & 27 & 6.69E-57\\
		3 & Endometrial cancer & 7 & 3.73E-15\\
		4 & Trans-sulfuration and one carbon metabolism & 6 & 2.02E-14\\
		5 & Antifolate resistance & 5 & 5.30E-12\\
		6 & Modulators of TCR signaling and T-cell activation & 5 & 2.34E-10\\
		7 & TP53 Network & 4 & 9.83E-10\\
		8 & Imatinib and chronic myeloid leukemia & 3 & 9.62E-07\\
		9 & Guanine / Thymine mispair binding & 2 & 0.00017\\
		\hline
	\end{tabular*}
\end{table*}

\begin{figure}[h!]
	\centering
	\includegraphics[width=3.5in]{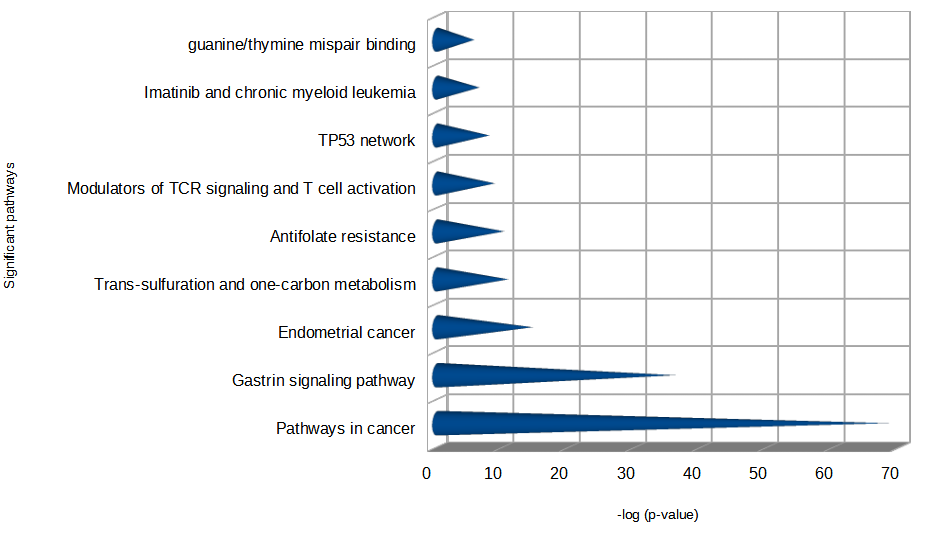}
	\caption{List of functionally enriched significant biological processes / pathways.}
	\label{N_C_D}
\end{figure}

\section{Results and Discussion}

The etiological mechanisms behind the oncogenesis caused by Epstein-Barr virus (EBV) infection in GC, NPC, HL and BL have been quite challenging, demanding the need for treatment strategies that utilize targeted drug-based therapies to understand the linkages between them. 

As recent studies suggest that EBV mainly affects B-lymphocytes along with NK-cells, T-cells and epithelial cells, it is pertinent to study and investigate the relation between GC, NPC, HL and BL among cancer patients \citep{hue2020epstein}. Our findings suggest that it is crucial to note the clinical manifestations of such carcinomas as they are inter-related based on modular detected communities involving critical biological processes involved in pathways in cancer, Imatinib and chronic myeloid leukemia and the TP53 network. Therefore, it becomes critical to investigate patients suffering from such conditions using a systemic approach based on over-representation analysis as it affects many critical biological processes common in EBV-associated carcinomas. Hence we implement community detection techniques that are useful for analyzing large-scale biological networks to discover key biological processes that overlap with each other \citep{rahiminejad2019topological}. We present insights obtained from network analysis of the disease-gene network and functional enrichment studies. 

\subsection{Disease-Gene Network Analysis}

Disease-gene network was constructed and the bipartite graph denoted edges connecting associated carcinomas linked with Epstein-Barr virus (mainly GC, NPC, HL and BL and few syndromes) to human genes. The network comprised of 404 nodes (382 human genes and 22 diseases) and 889 edges. 

The top 10 genes that were involved with EBV-associated carcinomas involving GC, NPC, HL and BL were identified as CASP10, BRAF, NFKBIA, IFNA2, GSTP1, CSF3, GATA3, UBR5, AXIN2 and POLE.  

\subsection{Network Topology Community Detection}

We performed community detection on the disease gene network (404 nodes and 889 edges) and applied the Louvain algorithm \citep{blondel2008fast} to obtain 18 significant communities. The p-value was kept less than 0.00001 to fetch significant communities. Further, we carried out functional enrichment to identify key biological processes in various carcinomas involved in GC, NPC, HL and BL. Figure 4 shows the network representation of top-ranked biological processes/pathways obtained through the analysis.

\subsection{Functional Enrichment Analysis using over-representation methodology}

We highlight the critical biological processes affected by the Epstein-Barr virus. The corresponding p-value threshold was found to be less than 2.43E-06, to identify significantly enriched biological processes. As listed in Table 3, we listed the community of genes involved with 9 key biological processes and critical pathways linked with the Epstein-Barr virus in the overall disease gene network. The significance level of the p-values was found to be in the range between 4.78E-69 to 2.43E-6. The 9 such key biological processes pathways (see Table 4) \label{tbl:biological_process} were found to be associated with the pathways in cancer, Gastrin signaling pathway, Endometrial cancer, Trans-sulfuration and one carbon metabolism, Antifolate resistance, Modulators of TCR signaling and T-cell activation, TP53 Network, Imatinib and chronic myeloid leukemia and Guanine / Thymine mispair binding. The significance of each of them is shown in Figure 5 \label{tbl:biological_process}.

\subsection{Validation}

For validation, we correlated our findings using the STRING database in the form of manually curated function protein interaction network database available for human proteins \citep{szklarczyk2021string}. As shown in Table 5 \label{tbl:fdr}, all 9 significant communities involving gene-ontologies were validated (supplementary table attached) with least False Discovery Rate (FDR), depicting that all of highlighted GO:BP terms are of statistical significance. 

\section{Conclusion}

In this study, we carried out over-representation analysis using community detection of the disease-gene network linked with Epstein-Barr virus (EBV) linked to carcinomas (mainly GC, NPC, HL and BL). We found that the above mentioned carcinomas are linked to a few common genes that are interconnected through various significant biological processes infected by Epstein-Barr virus. Further, it was observed that the ABL1 gene plays a vital role in 3 out of 9 biological processes based on over-representation analysis followed by other genes such as APC, BAX, BCR, BBC3, BAX, BIRC2, BIRC5, BRAF, BMP2, CCND1, CDK genes, MAPKs, PIK3CA and STAT3. Hence, the linkages between various biological processes infected with Epstein-Barr virus in causing carcinomas from the network of common genes present significant novel insights and therefore, we provide our case for clinical studies.

We report 9 key biological processes and disease pathways affected by Epstein-Barr virus (EBV) in the disease gene network. Using topological network analysis and community detection algorithm followed by functional enrichment, we identified the community of genes that were involved with respective biological processes involving carcinomas (GC, NPC, HL and BL). Since there are currently no specific treatments for EBV causing infectious mononucleosis other than the use of corticosteroids, further analysis of common biological processes and over-represented genes in the involvement of such carcinomas will provide better outcomes. Consequently, for EBV-associated carcinomas mainly GC, NPC, HL and BL, more such clinical investigations are needed for better prognostic and therapeutic outcomes.

\section*{Acknowledgments}

The authors would like to thank the Indian Institute of Information Technology, Allahabad (IIIT-A) for providing the infrastructural support and computational facilities.

%

\bibliographystyle{unsrt}
\bibliography{q-bio}   

\begin{thebibliography}{10}

\bibitem{epstein1964virus}
Michael~Anthony Epstein.
\newblock Virus particles in cultured lymphoblasts from burkitt's lymphoma.
\newblock {\em Lancet}, 1:702--703, 1964.

\bibitem{hue2020epstein}
Susan Swee-Shan Hue, Ming~Liang Oon, Shi Wang, Soo-Yong Tan, and Siok-Bian Ng.
\newblock Epstein--barr virus-associated t-and nk-cell lymphoproliferative
  diseases: an update and diagnostic approach.
\newblock {\em Pathology}, 52(1):111--127, 2020.

\bibitem{babcock1998ebv}
Gregory~J Babcock, Lisa~L Decker, Mark Volk, and David~A Thorley-Lawson.
\newblock Ebv persistence in memory b cells in vivo.
\newblock {\em Immunity}, 9(3):395--404, 1998.

\bibitem{henle1968relation}
Gertrude Henle, Werner Henle, and Volker Diehl.
\newblock Relation of burkitt's tumor-associated herpes-ytpe virus to
  infectious mononucleosis.
\newblock {\em Proceedings of the National Academy of Sciences}, 59(1):94--101,
  1968.

\bibitem{shannon2017epstein}
Claire Shannon-Lowe, Alan~B Rickinson, and Andrew~I Bell.
\newblock Epstein--barr virus-associated lymphomas.
\newblock {\em Philosophical Transactions of the Royal Society B: Biological
  Sciences}, 372(1732):20160271, 2017.

\bibitem{cohen2018vaccine}
Jeffrey~I Cohen.
\newblock Vaccine development for epstein-barr virus.
\newblock {\em Human Herpesviruses}, pages 477--493, 2018.

\bibitem{murata2014modes}
Takayuki Murata, Yoshitaka Sato, and Hiroshi Kimura.
\newblock Modes of infection and oncogenesis by the epstein--barr virus.
\newblock {\em Reviews in medical virology}, 24(4):242--253, 2014.

\bibitem{wang2021identification}
Zeyang Wang, Zhi Lv, Qian Xu, Liping Sun, and Yuan Yuan.
\newblock Identification of differential proteomics in epstein-barr
  virus-associated gastric cancer and related functional analysis.
\newblock {\em Cancer cell international}, 21(1):1--15, 2021.

\bibitem{farrell2019epstein}
Paul~J Farrell.
\newblock Epstein--barr virus and cancer.
\newblock {\em Annual Review of Pathology: Mechanisms of Disease}, 14:29--53,
  2019.

\bibitem{pinero2020disgenet}
Janet Pi{\~n}ero, Juan~Manuel Ram{\'\i}rez-Anguita, Josep Sa{\"u}ch-Pitarch,
  Francesco Ronzano, Emilio Centeno, Ferran Sanz, and Laura~I Furlong.
\newblock The disgenet knowledge platform for disease genomics: 2019 update.
\newblock {\em Nucleic acids research}, 48(D1):D845--D855, 2020.

\bibitem{thompson2004epstein}
Matthew~P Thompson and Razelle Kurzrock.
\newblock Epstein-barr virus and cancer.
\newblock {\em Clinical cancer research}, 10(3):803--821, 2004.

\bibitem{jeong2001lethality}
Hawoong Jeong, Sean~P Mason, A-L Barab{\'a}si, and Zoltan~N Oltvai.
\newblock Lethality and centrality in protein networks.
\newblock {\em Nature}, 411(6833):41--42, 2001.

\bibitem{brandes2008robert}
Gaertler Marco Gorke Robert Hoefer Martin Nikoloski~Zoran Brandes~Ulrik,
  Delling~Daniel and Wagner Dorothea.
\newblock On modularity clustering.
\newblock {\em IEEE Transactions on Knowledge and Data Engineering},
  20(2):172--188, 2008.

\bibitem{sham2014statistical}
Pak~C Sham and Shaun~M Purcell.
\newblock Statistical power and significance testing in large-scale genetic
  studies.
\newblock {\em Nature Reviews Genetics}, 15(5):335--346, 2014.

\bibitem{singhal2020multiscale}
Akshat Singhal, Song Cao, Christopher Churas, Dexter Pratt, Santo Fortunato,
  Fan Zheng, and Trey Ideker.
\newblock Multiscale community detection in cytoscape.
\newblock {\em PLoS computational biology}, 16(10):e1008239, 2020.

\bibitem{gulbahce2008art}
Natali Gulbahce and Sune Lehmann.
\newblock The art of community detection.
\newblock {\em BioEssays}, 30(10):934--938, 2008.

\bibitem{zheng2021hidef}
Fan Zheng, She Zhang, Christopher Churas, Dexter Pratt, Ivet Bahar, and Trey
  Ideker.
\newblock Hidef: identifying persistent structures in multiscale ‘omics data.
\newblock {\em Genome biology}, 22(1):1--15, 2021.

\bibitem{blondel2008fast}
Vincent~D Blondel, Jean-Loup Guillaume, Renaud Lambiotte, and Etienne Lefebvre.
\newblock Fast unfolding of communities in large networks.
\newblock {\em Journal of statistical mechanics: theory and experiment},
  2008(10):P10008, 2008.

\bibitem{raudvere2019g}
Uku Raudvere, Liis Kolberg, Ivan Kuzmin, Tambet Arak, Priit Adler, Hedi
  Peterson, and Jaak Vilo.
\newblock g: Profiler: a web server for functional enrichment analysis and
  conversions of gene lists (2019 update).
\newblock {\em Nucleic acids research}, 47(W1):W191--W198, 2019.

\bibitem{jaccard1912distribution}
Paul Jaccard.
\newblock The distribution of the flora in the alpine zone. 1.
\newblock {\em New phytologist}, 11(2):37--50, 1912.

\bibitem{storey2003statistical}
John~D Storey and Robert Tibshirani.
\newblock Statistical significance for genomewide studies.
\newblock {\em Proceedings of the National Academy of Sciences},
  100(16):9440--9445, 2003.

\bibitem{kanehisa2000kegg}
Minoru Kanehisa and Susumu Goto.
\newblock Kegg: kyoto encyclopedia of genes and genomes.
\newblock {\em Nucleic acids research}, 28(1):27--30, 2000.

\bibitem{pico2008wikipathways}
Alexander~R Pico, Thomas Kelder, Martijn~P Van~Iersel, Kristina Hanspers,
  Bruce~R Conklin, and Chris Evelo.
\newblock Wikipathways: pathway editing for the people.
\newblock {\em PLoS biology}, 6(7):e184, 2008.

\bibitem{gene2004gene}
Gene~Ontology Consortium.
\newblock The gene ontology (go) database and informatics resource.
\newblock {\em Nucleic acids research}, 32(suppl\_1):D258--D261, 2004.

\bibitem{rahiminejad2019topological}
Sara Rahiminejad, Mano~R Maurya, and Shankar Subramaniam.
\newblock Topological and functional comparison of community detection
  algorithms in biological networks.
\newblock {\em BMC bioinformatics}, 20(1):1--25, 2019.

\bibitem{szklarczyk2021string}
Damian Szklarczyk, Annika~L Gable, Katerina~C Nastou, David Lyon, Rebecca
  Kirsch, Sampo Pyysalo, Nadezhda~T Doncheva, Marc Legeay, Tao Fang, Peer Bork,
  et~al.
\newblock The string database in 2021: customizable protein--protein networks,
  and functional characterization of user-uploaded gene/measurement sets.
\newblock {\em Nucleic acids research}, 49(D1):D605--D612, 2021.

\end{thebibliography}

%
%


\end{document}